\documentclass{ws-ijtaf}

\usepackage{multirow}
\usepackage{color}

\begin{document}

\markboth{Basnarkov, Stojkoski, Utkovski and Kocarev}
{Option Pricing with Heavy-Tailed Distributions of Logarithmic Returns}

\catchline{}{}{}{}{}

\title{OPTION PRICING WITH HEAVY-TAILED DISTRIBUTIONS OF LOGARITHMIC RETURNS}

\author{LASKO BASNARKOV}
\address{Faculty of Computer Science and Engineering, SS. Cyril and Methodius University, Skopje, Macedonia\\
and\\
Macedonian Academy of Sciences and Arts, Skopje, Macedonia\\
\email{lasko.basnarkov@finki.ukim.mk} }

\author{VIKTOR STOJKOSKI}
\address{Macedonian Academy of Sciences and Arts, Skopje, Macedonia\\
\email{vstojkoski@manu.edu.mk} }

\author{ZORAN UTKOVSKI}
\address{Fraunhofer Heinrich Hertz Institute, Berlin, Germany\\
\email{zoran.utkovski@hhi.fraunhofer.de} }

\author{LJUPCO KOCAREV}
\address{Macedonian Academy of Sciences and Arts, Skopje, Macedonia\\
and\\
Faculty of Computer Science and Engineering, SS. Cyril and Methodius University, Skopje, Macedonia\\
\email{lkocarev@manu.edu.mk} }

\maketitle

\begin{history}
\received{(Day Month Year)}
\revised{(Day Month Year)}
\end{history}

\begin{abstract}
A growing body of literature suggests that heavy tailed distributions represent an adequate model for the observations of log returns of stocks. Motivated by these findings, here we develop a discrete time framework for pricing of European options. Probability density functions of log returns for different periods are conveniently taken to be convolutions of the Student's t-distribution with three degrees of freedom. The supports of these distributions are truncated in order to obtain finite values for the options. Within this framework, options with different strikes and maturities for one stock rely on a single parameter -- the standard deviation of the Student's t-distribution for unit period. We provide a study which shows that the distribution support width has weak influence on the option prices for certain range of values of the width. It is furthermore shown that such family of truncated distributions approximately satisfies the no-arbitrage principle and the put-call parity. The relevance of the pricing procedure is empirically verified by obtaining remarkably good match of the numerically computed values by our scheme to real market data.
\end{abstract}

\keywords{Asset pricing; Option pricing; Heavy-tailed distributions; Truncated distributions.}

\section{Introduction}

Options are transferable contracts between two parties that have spot prices which depend on the uncertain future price of the respective underlying asset. Consequently, a mixture of knowledge and intuition is needed in order to reach a good estimate for the so-called ``fair'' price. The mainstream tool used for theoretical calculation of option prices is the famous formula introduced in~\cite{black1973pricing, merton1973theory}. It relies on the assumption that the probability distribution for the log returns of the underlying asset is Gaussian. This distribution features mathematical convenience but is also omnipresent in nature, technology and society and is thus also known as ``normal''. For decades the scholars as well as the practitioners in different fields had little doubt in its relevance. However, one can find that both individuals within finance academic circles as well as option traders are familiar with the fact that the Gaussian underestimates the appearance of extreme movements of the stock prices. Details for this observation can be found for example in~\cite{mandelbrot1963variation, fama1965, plerou1999scaling, amaral2000distribution, haug2011option}. The more frequent than expected appearance of such events was even popularized among the general public in~\cite{taleb2007black}. Moreover, in~\cite{plerou1999scaling} and~\cite{amaral2000distribution} it was discovered that the tails of the distributions of returns seem to decay as an inverse of a power law. In order to overcome such observations, various modifications of the original Black-Scholes-Merton model have been developed, which can be roughly categorized in three groups. The followers of the first approach take the Black-Scholes-Merton formula as a basis and modify it by using other processes for modeling rare events, for example see~\cite{merton1976option, madan1998variance, kou2002jump, patel2018fourth}. Alternatively, \cite{cox1975notes, heston1993closed} and~\cite{hagan2002managing}, among others, use stochastic volatility for modeling the time variability of the price fluctuation intensity. Power-law-tailed distributions of returns are utilized as a starting point in the last group of models. Examples of such alternatives are given in  \cite{matacz2000,borland2002theory, borland2004non,moriconi2007} and \cite{cassidy2010pricing}. 
Another feature that favors the Gaussian distribution over others for being used as a building block in the modeling of logarithmic returns is that it is stable, i.e. a linear combination of independent random variables drawn from this distribution also has the same distribution. We note that, the L\'{e}vi distribution which is known to have power law tails, also has this property\footnote{The L\'{e}vy distributions are parameterized with a constant $0 < \alpha < 2$ and they are characterized with power law tails instead of the exponential decay of the Gaussian, which can be obtained for $\alpha = 2$. They are stable like the Gaussian which means that sum of two or more L\'{e}vy distributed variables is L\'{e}vy distributed as well. As one can see in \cite{bouchaud2003theory}, there are no other stable distributions besides them.}. Nevertheless, even though the tails of the histograms of observed log returns of stock prices seem to follow a power law, they are thinner than those of the L\'{e}vi kind, which implies that they fall off faster, and consequently possess finite variance. Concretely, in a detailed study of the historical data of price returns, in~\cite{plerou1999scaling, amaral2000distribution} it was shown that the power law tail index of the cumulative distribution is close to $\alpha = 3$, which is apart enough from the L\'{e}vy region $\alpha \in (0,2)$. As a consequence, a sum of random variables drawn from such distributions does not produce a variable distributed with the same kind -- no such distribution is stable. 

\cite{student1908probable} introduced a probability distribution applicable in statistics of small sample size, which today is known as the Student's t-distribution. It is parameterized with the number of degrees of freedom and is characterized with a power-law tail index $\alpha$. Although less tractable than the Gaussian, this distribution has also frequently appeared as a convenient alternative for describing the log returns of different assets. In particular, in~\cite{blattberg1974comparison} it was used for modeling stock dynamics, in~\cite{nadarajah2015note} for currencies, while \cite{platen2008empirical} have implemented it for modeling the returns of market indexes. Moreover, it was used for study of joint distribution by \cite{chicheportiche2012joint} and it was obtained that it can provide a good fit for strongly correlated stocks. Since this function fits the historical observations of the log returns rather well, particularly at the tails, it has already found implementations for theoretical pricing of options. In one non-Gaussian option valuation approach by~\cite{borland2002option,borland2002theory}, the Student's t-distribution was obtained as a result of a correlated stochastic process. In another proposal by~\cite{cassidy2010pricing}, as a basis was considered the theoretical result that a mixture of Gaussians with stochastic volatility, such that the inverse of the volatility is chi-squared distributed, follows the Student's t-distribution. Empirical verification of that fact by using returns of real data has provided justification for the same authors to apply it for options pricing.

In this paper, we also utilize the Student's t-distribution with three degrees of freedom as an appropriate distribution for describing the dynamics of the log returns of prices, and then apply it for pricing of European options. As such, this work can be compared to two earlier contributions in the option valuation theory. In the first approach by \cite{borland2002option,borland2002theory} and \cite{borland2004non}, the prices are assumed to be driven by stochastic processes with statistical feedback which results in distributions of log returns that are of Tsallis type -- a generalization of the Student's t-distribution. The resulting models for pricing options offer determination of the values for a whole range of strikes and maturities by using a single parameter. In the second contribution, \cite{cassidy2010pricing} and \cite{cassidy2013log} apply truncated Student's t-distribution directly, and provide an analysis for the position of the truncation point. Our valuation framework is primarily driven by the empirical observations rather than aiming to provide a strong analytical theory. We keep our focus as much as possible on the features of the returns of the stocks and the related options, and use well known and simple techniques to derive a convenient distribution. Nevertheless, we also provide an approximate analysis for the theoretical basis of the proposed pricing scheme.

In order to implement the Student's t-distribution for option pricing, its support needs to be truncated (i.e. its support has to be restricted in some finite domain), as otherwise the option prices would be infinite, see for e.g. \cite{bouchaud1994black, mccauley2007martingale} and \cite{cassidy2010pricing}. This implies that the set of extreme log returns far at the tails have to be neglected from the option price valuation. In practice, there are two reasons which are able to provide explanation for the implications created by this truncating procedure. First, in reality, it is safer to assume that stock price movements do not scale infinitely and have a rather unknown upper and lower limit. This follows from the observation that the dynamics of stock prices are driven by social behavior of individuals which are able to observe and base their investment decision on the available information for the underlying stock and market. In fact, a plethora of other phenomena exhibit similar properties, and therefore truncation has found application in a range of fields, spanning from environment issues \cite{maltamo2004estimation} to traffic \cite{cao2014modeling}. Second, while overly extreme events are possible in the non-truncated distribution, we show that there is a weak influence on the option prices for a certain range of values of the width. In particular, we find that there is a region with very slow increase of the option price as a function of distribution support width. In this aspect, the neglected part of the support extends to probabilities which correspond to extreme events happening rarer than once in a millennium. Therefore the option values obtained with the proposed procedure could be considered as stable enough. The stability practically implies that the price has low sensitivity to the location of the truncation point when it is inside a wide enough region. This further indicates that option prices obtained from distributions with truncated power law tails can be assumed to be fair.

We use the resulting truncated distribution as a building block for obtaining distributions that span different time intervals. The pricing scheme we propose here is based on discrete time and the distributions for longer periods are generated by convolutions of the elementary one. These distributions approximately provide arbitrage-free price dynamics and consequently can be used for pricing options. The predictive power of our model is compared to that of Black-Scholes-Merton and the one proposed by \cite{borland2002theory} as benchmarks. The latter comparison offers an additional dimension for the applicative power of our model since it uses the same distribution and can be used for options maturing at different dates.
When comparing the options prices obtained from our framework with those observed at the market, one could see that it shows good accuracy in estimating the market values of option prices with different strikes and expiry dates by using only one parameter. To remind the reader, the Black-Scholes-Merton formula needs a whole set of different volatilities to match the market values of options with different strikes and maturities. The weakness of our model is in its confinement to discrete time and that due to the calculations of the discrete Fourier transforms it has a limit on the horizon to which it can provide results, which we believe is more of numerical than of theoretical nature.

We remark that this valuation procedure is not dependent on the proposed distribution only and can be used with other distributions of log returns which can be, for example, some distributions that track more closely the observations. Such refinement could possibly result in better fit to the observed values from the option market.

The rest of the paper is organized as follows. In the next Section~\ref{sec:fair_price} we provide the general formula for calculation of the fair option price. After short introduction to insights of the price evolution in Section~\ref{sec:empirical_stock_prices}, in Section~\ref{sec:valuation_framework} we present the proposed valuation procedure, and elaborate in more details the issues which arise by implementing it. In Section~\ref{sec:data} we describe the data used for verification of our pricing method. Subsequently, in Section~\ref{sec:results} we present the results of application of our model on real data. The conclusion and discussion about future work are given in the final Section~\ref{sec:conclusions}.

\section{Fair price of European style option} \label{sec:fair_price}

A European option on a risky asset is a contract between two parties for possible future trade of that asset at certain strike price $K$ and at some date $T$. The owner of a ``call'' option has a right to buy, while the holder of a ``put'' option can sell the asset. The strike price $K$ and the expiration date $T$ are fixed and are determined in advance, at the moment of option writing, which for theoretical analysis one can safely assume to be $0$. European options can be exercised only at the moment of maturity $T$, and the owner of a call option would do so if the stock price has a market value $S(T)$ that is larger than the strike $K$. By buying the stock for $K$ and immediately selling it at the market for $S(T)$, the agent possessing the option would capitalize the difference $S(T)-K$. Otherwise, when $S(T) < K$, the option has value zero, since the stock can be purchased on the market for lower costs.
At the time of writing $0$, or any later moment $t$ before maturity, when the option is traded between an option holder to a new one, the current price of the underlying asset $S(t)$ is certain while there are only assumptions for its possible value at the maturity. The fair value of the option should be the one which does not favor neither the buyer, nor the seller. This means that, as seen from the current moment $t$, the option value at the maturity should equal the expected gain obtained by exercising the option. The expectation is considered as the mean value of all possible profits $S(T) - K$. We note that for calculation of this value and the current price of the option no-arbitrage arguments are applied. This assumption indicates that there exist risk-neutral probability distributions such that the expectation of the value of the asset at any future moment, and particularly at the maturity $T$, discounted by the riskless bond rate $r$, equals its price at the present moment $t$
\begin{equation}
    S(t) = e^{-r(T-t)} \mathbb{E}[S_T(\Omega)] = e^{-r(T-t)} \int_{\Omega} S_T(\Omega) p_T(\Omega) d\Omega.
\end{equation}
In the last equation, $\Omega$ is a random variable which models the noise due to stochastic events that determine the asset price, and $\mathbb{E}$ denotes the mathematical expectation. If alternatively, such probability measure $p_T$ does not exist, then arbitrage is possible. We point out that it is a rather challenging problem to have a good model at some previous moment $t$ of the probability distribution $p_T(\Omega)$ which will encompass all features that are typical for stock price dynamics and also one that does not allow arbitrage for different future moments. Once one has a risk-neutral probability that describes the future asset values, the current value of any contingency on that asset, and particularly an option, is set to exclude arbitrage on the option as well. As is given in \cite{ross2014introduction}, when the risk-neutral distribution of the future price of a stock is $p_T(S)$, the fair value of its European call option would be fair, if it is obtained with the same discounting on the expectation of its value at the maturity
\begin{equation}
C_t = e^{-r(T-t)} \mathbb{E}[C_T(\Omega)] = e^{-r(T-t)} \int_K^{\infty} (S - K) p_T(S) dS,
\label{eq:Call_price_from_stock}
\end{equation}
where the integral is calculated only for the region of prices where the option has non-vanishing value. Since the future is uncertain, we revert to expectations that some features or rules observed in the past will remain in the upcoming time. For example, any stock price is expected to grow on average, and consequently its distribution is about to change as time evolves, but its growth patterns are likely to persist. For that reason it is more convenient to directly model the probability distribution of the price changes instead of that of the prices. Knowing that prices grow exponentially on average, it is customary to express the relationship between the known current and the future random price as
\begin{equation}
S_T = S_t e^{\mu(T-t)+x},
\label{eq:price_relationships}
\end{equation}
where $\mu$ is the average stock growth rate, while $x$ is a random variable that models the excess price appreciation. This means that the difference of the logarithms of prices, or the log return will be modeled with some distribution 
\begin{equation}
p(x) = p\left[\log S_T - \log S_t- \mu(T-t)\right].
\label{eq:log_returns}
\end{equation}
In the last equation $p(x)$ is a shifted version of the distribution of log returns $p\left[\log S_T - \log S_t\right]$ for the amount of the drift $\mu(T-t)$. The distribution $p(x)$ plays a major role in the determination of the fair value of the options. Traditionally, the log returns are modeled rather conveniently with the Gaussian distribution. At one hand, this is appropriate whenever the random process is a result of many independent random forces, which would describe the price movement as a result of independent decisions made by the market participants. On the other hand, the Gaussian distribution has many favorable properties which make the solution more easily tractable than any other function. As such, this distribution is at the core of the famous Black-Scholes-Merton option pricing framework. In addition, the Gaussian distribution enables the construction of an alternative portfolio that results in the same value as the option, as well as providing easiness in the calculation of the other option attributes -- the Greeks. Refer to \cite{hull2017fundamentals} for more details about this topic.

To provide a specific example, we consider the formula for a European call option price on a stock with instantaneous return rate $\mu$ and riskless interest rate $r$, when the log return process is Gaussian
\begin{equation}
C_T(t) = e^{-r(T-t)}\int_{\log K/S_t - \mu(T-t)}^{\infty} [S_t e^{\mu(T-t) + x} - K] \frac{1}{\sqrt{2\pi \sigma}} e^{-\frac{x^2}{2\sigma^2}} dx.
\label{eq:BSM_Call}
\end{equation}
The lower integration bound is obtained from the situation when the stock price at maturity equals the strike
\begin{equation}
S_t e^{\mu(T-t) + x_l} = K.
\end{equation}
One can notice that the convergence of the integral (\ref{eq:BSM_Call}) is ensured by faster decay of the Gaussian distribution as compared to the exponential price growth. We note that when the log returns follow the Gaussian, the distribution of the price will be risk-neutral if one chooses $r = \mu + \sigma^2/2$. Therefore, as is shown in \cite{ross2014introduction}, one can obtain a much simpler version of the formula, and close form expressions for the option price.

When one uses a distribution with fat tails, the option pricing integral (\ref{eq:Call_price_from_stock}) would diverge since the exponential price term cannot be compensated with any power law decay. \cite{bouchaud1994black}, \cite{cassidy2013log} and \cite{mccauley2007martingale} provide explanation of  such a problem. In this case, one has to find some remedy, such as truncating the distribution sharply like \cite{cassidy2010pricing} have suggested, or using one with far part of the tail falling off exponentially as is proposed by \cite{moriconi2007} and \cite{cassidy2012effective}. Clearly, the way of modeling such heavy tailed distribution should be supported by theoretical explanation or by empirical evidence. However, when one does not have sufficient data for precise modeling of the extreme price shocks, the estimation of the tail of the distribution of log returns is mixture between guesswork and convenience. Our choice was based on convenience, without neglecting the empirical relevance of the selected probability distribution. Due to the simplicity we have chosen the first approach, which means that the call option price would be calculated from 
\begin{equation}
C_T(t) = e^{-r(T-t)} \int_{\log K/S_t - \mu(T-t)} ^{x_{\max}} [S_t e^{\mu(T-t) + x} - K] p_T(x) dx,
\label{eq:Call_from_price_chng}
\end{equation} 
where $x_{\max}$ is the point where the distribution is truncated. Once one has fully determined the option price formula (\ref{eq:Call_from_price_chng}), when Student's t-distribution is used for log return, the Greeks can be easily calculated as was shown by \cite{cassidy2013log}. Note that the same convergence problem would not appear for put options, which are characterized with nonzero value at maturity when the stock price $S$ is lower than the strike $K$. Hence, by averaging over the distribution of the returns on the underlying stock the current put option price is
\begin{equation}
P_T(t) = e^{-r(T-t)}\int_{-\infty}^{\log K/S_t - \mu(T-t)} [K - S_t e^{\mu(T-t) + x}] p_T(x) dx.
\label{eq:Put_from_price_chng}
\end{equation}
The last integral converges due to the exponential decay of the price at the first limit of the integration regardless of the decay type of the distribution. However, as we will see later, the prices for the put options can be calculated from those of the calls, by using the well known put-call parity. The reader can find more details about this relationship in \cite{hull2017fundamentals}.

\section{Empirical observations of stock price dynamics} \label{sec:empirical_stock_prices}

A view on stock price charts of any company traded at developed market would not reveal any regular mechanism causing its changes. It features random wiggling even on smallest time scales. There is no doubt that any model of such financial time series should include certain level of randomness and involve appropriate probability distributions. However, to our knowledge, there is no probability distribution of returns that incorporates all observed features while providing easy mathematical tractability. 

The Gaussian or the normal distribution is the one that should be expected whenever the variable under observation is influenced by a sum of many independent random forces with identical variance. Stock price changes are result of the demands and offers of the market actors which have different needs and opinions about the stock's worth. It is a plausible assumption that their decisions are more or less independent and consequently their behavior would drive the price in such a way that its changes within any interval of observation will be normally distributed. Moreover, due to the very well developed mathematical tools for the Gaussian, one is pushed towards believing that this distribution is the most appropriate one.

However, the studies of the histograms of log returns of prices do not correspond to the Gaussian distribution when one accounts for extreme events. They seem to have rather fat tails, which practically means that large changes, whether price appreciations or crushes, happen more often than the Gaussian distribution predicts. In a particular study about the cotton prices made about half a century ago, \cite{mandelbrot1963variation} has suggested that the corresponding changes could be more appropriately modeled with L\'{e}vy distribution with tail index $\alpha = 1.7$, which means that the corresponding probability density falls off as $1 / x^{2.7}$ function. More recently, in a wider study encompassing much more data from different stocks, \cite{plerou1999scaling} and \cite{amaral2000distribution} have obtained that for short periods of observation the price changes distributions indeed fall off as power law but much faster, characterized with tail index around 3 $(\alpha \approx 3)$. This is apart enough from the L\'{e}vy region (0,2) which means that these distribution functions have finite variance. According to the Central Limit Theorem, the sum of independent random variables drawn from identical distributions with finite variance converges towards the normal distribution when the number of the variables grows indefinitely. Then one should not be surprised from the findings within the same work which show that for longer periods the price changes seem to converge slowly towards Gaussian distribution. The convergence rate towards Gaussian is very nicely explained in \cite{bouchaud2003theory}. Thus, one is suggested to believe that an appropriate model of price fluctuations should apply fat tail distributions for shorter periods, while the Gaussian should be used when one considers returns for longer periods, as is done, for instance, by \cite{cassidy2011describing}. We note that there have been attempts for theoretical explanation of the emergence of heavy tails. In specific examples, \cite{lux1999scaling} use the interaction between the agents as a reason to prove the appearance of scaling, whereas \cite{gabaix2003theory} attribute these phenomena to the action of large market players.

\section{Option valuation framework} \label{sec:valuation_framework}

\subsection{Mathematical background}
\label{sec:math_background}

Here we provide a short overview of the basic mathematical tools needed to understand the full potential of the proposed pricing framework. In a discrete time system as respective time unit is usually taken the smallest interval at which changes of the system happen. At the end of each such interval $i$ the price changes its current value $S_i$ to $S_{i+1} = X_{i+1} S_i$, where $X_{i+1}$ is a random multiplier which encapsulates the change. The multipliers are usually taken to be independent, and for simplicity, one can consider a scenario where the initial price is normalized to $S_0 = 1$. If at every interval, the price is multiplied by some random number $X_i$, then its random value at moment $N$ in the future would be
\begin{equation}
S_N = S_0 \prod_{i=1}^N X_i = \prod_{i=1}^N X_i.
\label{eq:final_price}
\end{equation}
Since, on average, all prices grow, it is convenient to extract the mean growth factor and express the random multiplier as
\begin{equation}
X_i = M \tilde{X}_i= e^{\mu+x_i},
\end{equation}
where $\mu$ is the mean growth rate 
$\mu = \log M$ and $x_i = \log \tilde{X}_i$ is the random excess growth rate. Assuming that the mean growth rate is constant, the future stock price can be expressed as
\begin{equation}
S_N = e^{\mu N} \prod_{i=1}^N e^{x_i}.
\end{equation}
The last expression suggests that in order to obtain the probability distribution of the final price $S_N$ one should find a distribution of a product of random variables. When one multiplies random numbers instead of adding them, in order to apply the mathematical machinery available, it is more appropriate to consider the distribution of their logarithms, which results in the logarithm of the price given as
\begin{equation}
\log S_N = \mu N + \sum_{i=1}^N x_i.
\label{eq:final_price_log}
\end{equation}
Then, from a known result in probability theory, the probability density function we are searching for, is obtained from convolution of the probability densities of the logarithms of the individual random multipliers. More precisely, when each multiplier has respective density $p_i(x) = \mathbf{Prob}(\log \tilde{X}_i=x)$ the distribution of the logarithm of the final price is given by the convolution
\begin{equation}
\mathbf{Prob}(\log S_N - \mu N = x) = (p_1 * p_2 * \cdots * p_N)(x),
\label{eq:price_convolution}
\end{equation}
where the convolution of two functions is the integral
\begin{equation}
(f * g) (x) = \int_{-\infty}^{\infty} f(y) g(x - y) dy. 
\label{eq:convolution_definition}
\end{equation}
The convolution operation has a very useful property which states that a Fourier transform of convolution of two functions is product of the Fourier transforms of those functions. In the finance literature, one can find about the convolution operation for example in \cite{bouchaud2003theory}. Then, the Fourier transform of the probability of the logarithm of the final price is
\begin{equation}
\mathrm{p}_N = \mathcal{F}[\mathbf{Prob}(\log S_N - \mu N = x)] = \prod_{i=1}^N \mathrm{p}_i,
\label{eq:Fourier_tranform}
\end{equation}
where
\begin{equation}
\mathrm{p}_i = \mathcal{F}[p_i(x)]
\end{equation}
is the Fourier transform of the distribution of the logarithm of the individual price increment at the iteration $i$. Going back to the original problem, the probability distribution of the log price is obtained with inverse Fourier transform
\begin{equation}
P_N (x) = \mathbf{Prob}(\log S_N - \mu N = x)= \mathcal{F}^{-1}\left[\prod_{i=1}^N \mathcal{F}[p_i(x)] \right].
\end{equation}
The last formula can be used for any distribution of logarithm of price changes (log returns) for which the Fourier transform exists. For example, the Gaussian distribution is very attractive for such applications since its Fourier transform has also exponential form, and consequently, a sum of any number of Gaussian distributed numbers has the same distribution. 

\subsection{Modeling probability distributions of returns}
\label{sec:returns_model}

A direct approach for selecting an adequate probability distribution for the price returns is to use empirical distributions obtained by fitting historical data of stock prices. Nevertheless, one should be cautious of the reliability of these distributions due to the lack of sufficient data which is necessary for good fitting, especially at the tips of the tails. Another path that can be followed involves using distribution functions which have desirable properties such as power law tails, as is observed from the stock markets, or have known Fourier transforms with a possibly simple enough form and which offer tractable mathematical analysis, or something similar which results in better convenience in terms of analytical treatment. 

In this regard, even earlier studies by \cite{fama1965} and \cite{blattberg1974comparison} have empirically shown that the tails of the distribution of observed stock log returns are better modeled if they are taken to fall off as a power law instead of exponentially as Gaussian distribution does. Moreover, as is shown in \cite{plerou1999scaling} and \cite{amaral2000distribution}, the tail index of the cumulative distribution is close to three, which means that the corresponding probability density falls off as $1 / x^4$. A probability density function with such tail index is the Student's t-distribution with three degrees of freedom\footnote{Note that the Tsallis distributions constitute a wider class with real number tail indices making the Student's t-distribution a special case of it. See \cite{queiros2005power} for more insight about the Tsallis distribution.}. Consequently, many works, like those by \cite{borland2002theory,borland2004non} and \cite{cassidy2010pricing}, that propose option valuation formulas based on the Student's and related distributions have started to emerge. The Student's t-distribution is identified with the number of degrees of freedom and a width parameter $\gamma$. The density of this distribution with three degrees of freedom is
\begin{equation}
p_{S}(x) = \frac{2 \gamma^3} {\pi(\gamma^2 + x^2)^2}.
\end{equation}
It is also very convenient, because it has rather simple Fourier transform
\begin{equation}
F_{S}(\omega) = \frac{1}{\sqrt{2\pi}} (1 + \gamma|\omega|) e^{-\gamma|\omega|}.
\label{eq:Four_Student}
\end{equation}
It is worth noting that the standard deviation $\sigma = \sqrt{\mathbb{E}[x^2 p_{S}(x)]}$ of this case equals its parameter $\gamma = \sigma$. By the convolution property (\ref{eq:Fourier_tranform}), a sum of $N$ random variables drawn from this distribution has a Fourier transform which is a power of the last function, i.e.,
\begin{equation}
F_{G,N}(\omega) = \frac{1}{\sqrt{2\pi}} (1 + \gamma|\omega|)^N e^{-N\gamma|\omega|}.
\label{eq:Four_Student_N}
\end{equation}
The distribution of the sum of many one step log returns is obtained with the inverse Fourier transform. By expanding the Fourier transform for small argument around the origin, as is shown in \cite{bouchaud2003theory}, the convolutions of any order have the same tail index. Because it has finite variance, accordingly to the Central Limit Theorem, a sum of infinitely many random variables drawn from this distribution follows a Gaussian distribution. However, the convergence toward Gaussian is such that the convolution of a large number of elementary distributions is Gaussian-like only at the central region around the mean, but not the tails. As seen from another point of view, the more summands one has (larger $N$) the tail dominance region is pushed further towards infinity, and thus the sum becomes more like the normal distribution mainly at the body and upper part of the tails. An analysis of the potential of these features of the Student's t-distributions for option valuation has been discussed in \cite{cassidy2011describing}.

\subsection{Truncation of heavy-tailed distributions of returns}

In developing a theoretical valuation framework one needs distributions of the logarithmic stock returns for different periods of observations in order to be able to price the options with different maturities. For the stable distributions like the Gaussian and L\'{e}vy ones, convolution of any number of variables is distributed like one variable up to a scale factor. Thus, one needs to estimate only the appropriate parameter of the distribution for certain period and then automatically has them completely determined for any period. For other distribution types the functional form is not preserved under the convolution operation. Thus one cannot use a family of Student's t-distributions with different variances for modeling log returns for different periods if they represent sums of independent and identically distributed random variables. Although the fall-off parameter at the tails is preserved under convolution, the body of the distribution starts to resemble the Gaussian as one adds additional variables. This reasoning leads to the idea that one should try to take the Student's t-distribution with appropriately fit variance for certain unit period and then use its convolutions for multiple periods. 

Unfortunately, as it was already seen, the power law decay cannot compensate the exponential growth of the price in calculation of a call option value, and one must either change the tip of the tail of the distribution to exponential or truncate it as given in equation (\ref{eq:Call_from_price_chng}). A truncated version $f_{\mathrm{trunc}}(x)$ of a function with infinite support is one that is identical to the original inside some region $(x_{\min},x_{\max})$ while zero outside
\begin{equation}
f_{\mathrm{trunc}}(x) = \left\{ \begin{array}{ll}
		f(x)  & x_{\min} \geq x \geq x_{\max}; \\
		0 & \mathrm{elsewhere}.
	\end{array}
\right.
\end{equation}
Any truncated function can be represented as a product of the non-truncated version and a rectangle function
\begin{equation}
\mathrm{rect}(x) = \left\{ \begin{array}{ll}
		1  & x_{\min} \geq x \geq x_{\max}; \\
		0 & \mathrm{elsewhere},
	\end{array}
\right.
\end{equation}
which has value one only in the part that remains from truncation, while zero outside. Then, the characteristic function of the truncated function, as a product of two functions, is convolution of the original and that of the rectangle due to the convolution property of the Fourier transform. The Fourier transform of the rectangle is the Sinc function $\mathrm{sinc}(x) = \sin (x)/x$ which is also a spherical Bessel function of order zero. In this case, the Fourier transform of the truncated distribution is calculated as an integral of the form 
(see eq. (\ref{eq:convolution_definition}))
\begin{equation}
   \mathcal{F}_{\mathrm{trunc}}(\omega) = \int_{-\infty}^{\infty}  \mathcal{F}(\eta) \frac{\sin (\omega - \eta)} {\omega - \eta} d\eta.
   \label{eq:trunc_S_Fourier}
\end{equation}
For the case of the Student's t-distribution, the Fourier transform we need is an integral of product of exponential function with real argument and shifted spherical Bessel function. In the literature, to the best of our knowledge, there is no closed form solution for such integration. This implies that the Fourier transform of the truncated Student's t-distribution cannot be expressed through the known elementary or special functions. To remind, that is the building block of our family of distributions, because the price change distributions for various intervals are inverse Fourier transforms of powers of functions like (\ref{eq:trunc_S_Fourier}). Thus, if someone is willing to pursue along this path, one must rely only on numerical methods. The rest of the recipe would include numerical determination of the convolution (\ref{eq:trunc_S_Fourier}) of the Fourier transform of the Student's t-distribution with the Sinc function, then raising it on desired power and calculating inverse Fourier transform. In order to obtain the probability density one should finally make appropriate rescaling due to the probability loss which resulted from truncation. 

\begin{figure}[t!]
\begin{center}
\includegraphics[width=8cm]{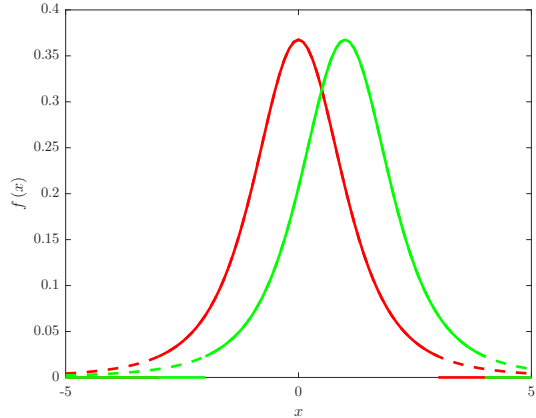}
\caption{A sketch of two Student's t-distributions (dashed) with their truncated versions (full). Note that the truncated and non-truncated functions coincide, where the first one is nonzero. This coincidence happens when the truncated function is not normalized.
\label{fig:trunc-distribution}}
\end{center}
\end{figure}

In another, approximate, but more simple approach, one can first obtain the necessary number of convolutions of the elementary Student's t-distribution and then make the truncation. In this way one could capture the tail behavior, while simultaneously providing Gaussian-like body of the distribution at the limit. To remind, the extensive analysis of the stock price returns by \cite{plerou1999scaling} and \cite{amaral2000distribution} have suggested power law tails for smaller periods and resemblance to Gaussian for longer periods. We emphasize that this approach provides functions which are only approximation of convolutions of truncated Student's t-distributions. For reasons that will be explained later in Section~\ref{sec:plateau}, here we chose to also truncate the convolutions in the same interval $(-M\gamma, M\gamma)$ and accordingly produce an analysis of the quality of the approximation only in that interval. In order to understand this better, one should first observe that the total probability corresponding to rare events, which are removed by the truncation and that are outside of central region of the Student's t-distribution for large number $M$ of standard deviations is bounded as
\begin{equation}
P_{\mathrm{extreme}}=2\int_{M\gamma}^{\infty} \frac{2 \gamma^3} {\pi[\gamma^2 + x^2]^2}dx < \frac{4 \gamma^3}{\pi} \int_{M\gamma}^{\infty} \frac{dx}{x^4} = \frac{4}{3\pi M^3}.
\label{eq:P_extreme}
\end{equation}
This means that the Student's t-distribution which is truncated far at the tails remains nearly normalized. Similar result should hold for the convolutions of it. Denote by $p_{S}^{(2)}$ the convolution of two Student's t-distributions and its truncated version by $p_{S,T}^{(2)}$. Also, let $p_{TS}^{(2)}$ is the convolution of two truncated Student's t-distributions. For clarification of this reasoning in figure \ref{fig:trunc-distribution} are shown two shifted Student's distributions with dashed curves and their (not normalized) truncated versions with full curves. By definition, convolution at each point is an integral of the product of the two functions. Thus the convolution of two truncated Student's distributions will miss the part of the integral where any of the functions is zero.

\begin{figure}[t!]
\begin{center}
\includegraphics[width=12cm]{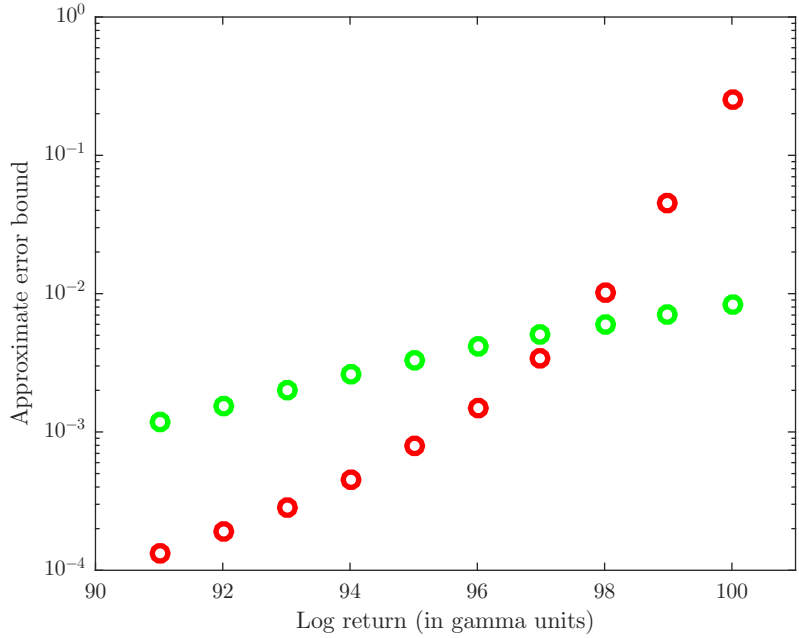}
\caption{Bounds on the difference between truncating convolutions of Student's t-distributions and convolutions of truncated Student's t-distributions. The red circles correspond to 2 distributions, while the green ones for convolution of 50 distributions. In this example the Student's t-distribution has $\gamma = 0.02$ and the truncation point is at 100 standard deviations, or $M\gamma = 2$.
\label{fig:Conv_err}}
\end{center}
\end{figure}

The difference between the two convolutions $\epsilon_2(y) = p_{S,T}^{(2)}(y) - p_{TS}^{(2)}(y)$ is given by two integrals of the following form (refer to Figure \ref{fig:trunc-distribution})
\begin{equation}
\epsilon_2(y) \approx 2\int_{M\gamma}^{\infty} p_S(x) p_S(x-y)dx = 2\int_{M\gamma}^{\infty} \frac{2 \gamma^3} {\pi(\gamma^2 + x^2)^2} \frac{2 \gamma^3} {\pi[\gamma^2 + (x-y)^2]^2}dx,
\label{eq:conv_error_2}
\end{equation}
where $y$ is the point where we look at the error, or the displacement between the peaks of the two distributions involved in the integral. We remind that the error is approximation because the truncated distributions should be normalized, and we have used the symmetry of the distribution. As one can see from \cite{hardy1952inequalities} a good estimate of the error can be obtained by using the H\"{o}lder inequality for integrals 
\begin{equation}
\int |f(x) g(x)| dx \leq \left(\int|f(x)|^\mathrm{p} dx\right)^{1/\mathrm{p}} \left(\int|f(x)|^\mathrm{q} dx\right)^{1/\mathrm{q}},
\end{equation}
which holds for real powers $\mathrm{p}, \mathrm{q} \in [1, \infty)$ satisfying $1/\mathrm{p}+1/\mathrm{q}=1$. Then the estimation of the error reads
\begin{eqnarray}
\epsilon_2(y) &\leq& 2\left[\int_{M\gamma}^{\infty}\left(\frac{2 \gamma^3} {\pi(\gamma^2 + x^2)^2}\right)^\mathrm{p} dx\right]^{1/\mathrm{p}} \left[\int_{M\gamma}^{\infty}\left(\frac{2 \gamma^3} {\pi[\gamma^2 + (x-y)^2]^2}\right)^\mathrm{q} dx\right]^{1/\mathrm{q}}\nonumber\\
&=&
2\left[\int_{M\gamma}^{\infty}\left(\frac{2 \gamma^3} {\pi(\gamma^2 + x^2)^2}\right)^\mathrm{p} dx\right]^{1/\mathrm{p}} \left[\int_{M\gamma-y}^{\infty}\left(\frac{2 \gamma^3} {\pi(\gamma^2 + x^2)^2}\right)^\mathrm{q} dx\right]^{1/\mathrm{q}}.
\end{eqnarray}
One can easily notice from the second integral in the last expression that the error grows with $y$ reaching its maximum at $y=M\gamma$. Since the integrand in the first integral decays faster, it is more convenient to take $\mathrm{p}$ to be larger than $\mathrm{q}$. By using the following inequality
\begin{equation}
\frac{1}{(\gamma^2 + x^2)^2} < \frac{1}{x^4},
\end{equation}
for $\mathrm{p} \to \infty$ one can obtain the following bound for the first term in the error estimation 

\begin{equation}
\left[\int_{M\gamma}^{\infty}\left(\frac{2 \gamma^3} {\pi(\gamma^2 + x^2)^2}\right)^\mathrm{p} dx\right]^{1/\mathrm{p}} < \frac{2\gamma^3}{\pi} \left[\int_{M\gamma}^{\infty} \frac{dx}{x^{4\mathrm{p}}} \right]^{1/\mathrm{p}} = \frac{2\gamma^3 (M\gamma)^{\frac{1-4\mathrm{p}}{\mathrm{p}}}}{\pi (4\mathrm{p}-1)^{\frac{1}{\mathrm{p}}}}\approx \frac{2}{\pi \gamma M^4}. \label{eq:conv_err_1}
\end{equation}
Due to the relationship between the parameters $p$ and $q$ in the H\"{o}lder inequality, for $p\to\infty$ one should take $q=1$, or 
\begin{equation}
\left[\int_{M\gamma-y}^{\infty}\left(\frac{2 \gamma^3} {\pi(\gamma^2 + x^2)^2}\right)^q dx\right]^{1/q} = \int_{M\gamma-y}^{\infty}\frac{2 \gamma^3} {\pi(\gamma^2 + x^2)^2} dx. \label{eq:conv_err_2}
\end{equation}
This integral has its maximal value $1/2$ when $y=M\gamma$ due to the normalization, while for general values of $y$ it is smaller. In figure \ref{fig:Conv_err} with red curve is given the estimate of the bound of the error by using (\ref{eq:conv_err_1}) and (\ref{eq:conv_err_2}) and divided with the probability density of the convolution at that point $\epsilon(y) / p_{S}^{(2)} (y)$. One can see that the error becomes significant only when $y$ is few standard deviations from the truncation point $M\gamma$. It means that instead of convolution of two truncated Student's t-distributions, within the region of interest, $(-M\gamma, M\gamma)$, one can take truncation of convolution of two Student's t-distributions. By using induction one can obtain that such estimate of the error can be obtained for convolutions of many Student's t-distributions as well. For that reason, denote in the same manner as above the convolution of $n-1$ Student's t-distributions as $p_{S}^{(n-1)}$.
Now, take as an induction hypothesis that truncating convolution of $n-1$ distributions $p_{S,T}^{(n-1)}$ is an approximation of convolution of $n-1$ truncated distributions $p_{TS}^{(n-1)}$. Denote by $p_{TS}^{(n-1),1}$ the result of convolution of this approximation with one truncated Student's t-distribution $p_{TS}$. This should be an approximation of convolution of $n$ truncated Student's t-distributions $p_{TS}^{(n)}$. Then the error between the truncated convolution of $n$ distributions $p_{S,T}^{(n)}$ and the convolution of $n$ truncated distributions $p_{TS}^{(n)}$ will be
\begin{equation}
\epsilon_n(y) = p_{S,T}^{(n)} - p_{TS}^{(n)} \approx p_{S,T}^{(n)} - p_{TS}^{(n-1),1} = p_{S,T}^{(n)} - p_{S,T}^{(n-1)} * p_{S,T},
\end{equation}
where we remind that the star denotes the convolution operation. The last approximation will be calculated by integral similar to (\ref{eq:conv_error_2}) which is a product of the tails of the Student's t-distribution and a convolution of $n-1$ functions of that kind. Consequently, the goodness of this approximation $\epsilon_n (y) = $ by the by H\"{o}lder inequality will be bounded as
\begin{equation}
\epsilon_n (y)\leq 2\left[\int_{M\gamma}^{\infty}\left(\frac{2 \gamma^3} {\pi(\gamma^2 + x^2)^2}\right)^p dx\right]^{1/p} \left[\int_{M\gamma - y}^{\infty}\left[p_{S}^{(n-1)}(x)\right]^q dx\right]^{1/q}.
\end{equation}
Again, for $p\to\infty$ the first term in the last equation is bounded as in~(\ref{eq:conv_err_1}). The second can be studied numerically by using $n-1$ convolutions of the Student's t-distribution. However, because the closed form of these is not known, we have obtained them from their Fourier transforms. As will be explained in more details below, the Fourier transform and its inverse are conveniently studied numerically by Fast Fourier Algorithm, which provides samples of the functions. For this purpose, we have used samples in the truncated region only, $(-M\gamma, M\gamma)$, since for up to several dozens of convolutions of the Student's t-distribution, most of the probability mass is contained there, which was verified numerically. This means that the integration only within this region does not differ much from the proper one which stretches to infinity. In figure~\ref{fig:Conv_err} is given a bound of the error between truncating $n$ convolutions $p_{S,T}^{(n)}$ and convolution of truncated Student's t-distribution $p_{TS}$ and truncated convolution of $n-1$ distributions $p_{S,T}^{(n-1)}$, divided by the convolution of $n$ distributions at every point of interest. As can be seen, such relative error in the probability density is non vanishing only near the truncation point. This might suggest that one could make significant error in estimating the extreme events, which is probably correct. But, our knowledge of the distributions of log returns in this region is very rough due to the insufficiency of data, so we are also not sure that the distribution has such tails in that region as well. We could summarize that, for practical reasons, one can use convolutions of the Student's t-distribution and then perform truncation in order to obtain distributions of log returns for different horizons.

\subsection{No-Arbitrage Principle}

One of the key concepts in the finance is the \emph{absence of free lunch}. This is a simplified expression which implies that if there is some mispriced item in the market, the traders would surely notice it and immediately start buying or selling depending on the direction of misprice and their actions would push the price towards the proper value. Nowadays, when information is spread rather quickly, one expects such arbitrage opportunities to not exist, apart on very short periods. One of the most important mathematical results in finance is the fundamental theorem of asset pricing. See \cite{harrison1981martingales}, \cite{kreps1981arbitrage} and \cite{delbaen1994general} for proofs of its variants. It states that the absence of arbitrage in a market coincides to the existence of equivalent risk neutral probability measure. This probability measure is different from the one observed in the real world. The expectation of the future value of an asset with respect to the real observations usually produces larger value, as compared to that obtained when the neutral probability is used. This can be explained with the fact that investment in a stock is riskier than in a bond and the investor in the former must be compensated for that. A recent work by \cite{cassidy2018risk} provides an engaging analysis when the these two probabilities differ. This theoretical measure is very useful, since it allows determination of the fair values of spot prices of stocks and their derivatives. The price of a derivative is the expectation of its apparently random future value calculated with respect to the risk neutral probability and then discounted back to any previous time, and particularly to the current moment. The discounting is performed with the riskless bond rate $r$. By the arbitrage theorem, this indicates that an investor can not make a sure profit out of nothing, or alternatively, no portfolio can grow more than the rate $r$ without risk. To be more specific, taking that the expected value of a stock in some future moment $T$ is $\mathbb{E}[S(T)]$, discounting it with the bank rate should result in its present value
\begin{equation}
S(t) = \mathbb{E}[S(T)e^{-r(T-t)}].
\end{equation}
As expressed in terms of the fundamental theorem of asset pricing, the process modeling the price of a stock will be appropriate if knowing its value $S(u)$ in the past up to the present moment $t$, provides the following estimate of the average of its future development
\begin{equation}
\mathbb{E}_P[e^{-rT}S(T)|S(u),0\leq u\leq t] = e^{-rt}S(t),
\end{equation}
where the average is calculated with respect to the risk neutral probability density $P = p(x)$. The last expression can be found in \cite{ross2014introduction}. If for a stock such distribution exists, than, by the theorem, the same distribution can be used for obtaining fair values of contingencies on that stock. Thus, when one tries to use some probability distribution for such purpose, one should first verify whether it is in accordance to the theorem. 

In this work we model the price dynamics in discrete time and thus any future price is given as
\begin{equation}
S(t) = S(0)e^{\mu t + x},
\end{equation}
where $\mu$ is the single step rate of return of the stock, while $x$ is the random excess return, which due to the numerous historical observations of the stock markets, we will conveniently describe its probability with truncated $t-$ fold convolutions of Student's t-distribution with three degrees of freedom. The expectation taken with respect to the truncated convolutions of the Student's t-distribution will be
\begin{equation}
\mathbb{E}_{P_{T}}[e^{-rT}S(T)|S(t)] = e^{-rT}\int_{-x_{\max}}^{x_{\max}}S(t) e^{\mu (T-t)+x} p_T(x) dx.
\label{eq:Student_No_Arbitrage}
\end{equation}
In the last equation we have intentionally put index $T$ to the risk neutral distribution $P$ in order to emphasize that it is different for different moments $T$. One can notice that in the last equation appears the moment generating function of the random log return
\begin{equation}
\int_{-x_{\max}}^{x_{\max}}e^x p_T(x) dx  = \mathbb{E}[e^{X}] = \sum_{i=0}^{\infty}\frac{m^i}{i!},
\end{equation}
where $m_i$ is the $i$-th moment. For symmetrical distributions only even order moments are nonzero. Also, when the higher order moments can be neglected one can obtain simpler approximation. In general, it is not easy to determine all moments for different periods $T$. However, for independent and identically distributed variables, the second moment accumulates linearly with time, so within the period $T-t$ one has
\begin{equation}
\int_{-x_{\max}}^{x_{\max}} x^2 p_T(x) dx = (T-t)\gamma^2,
\end{equation}
where $\gamma^2$ is the variance corresponding to single step. When the higher order moments can be neglected, which for example is the case when the variance is rather small $\gamma^2 \ll 1$, one has 
\begin{equation}
\int_{-x_{\max}}^{x_{\max}}e^{\mu (T-t)+x} p_T(x) dx \approx e^{\mu (T-t)} [1 + \frac{(T-t)\gamma^2}{2}] \approx e^{(\mu + \gamma^2 / 2) (T-t)}.
\label{eq:no_arbitrage}
\end{equation}
Like the case with the Brownian motion, one could ask for similar relationship between the bank rate, stock's growth rate and the variance of the distribution 
\begin{equation}
r = \mu + \gamma^2 / 2.
\end{equation}
This would imply that the convolutions of the truncated Student's distribution are approximately risk neutral distributions in the case of discrete time dynamics
\begin{equation}
\mathbb{E}_{P_T}[e^{-rT}S(T)|S(t)] \approx e^{-rt}S(t).
\end{equation}
In order to estimate the accuracy of the last expression we have numerically calculated the following difference \begin{equation}
\int_{-x_{\max}}^{x_{\max}}e^x p_T(x) dx - e^{T\gamma^2 / 2},
\end{equation}
for various moments $T$ for truncated Student's t-distribution with daily standard deviation $\gamma = 0.02$. It was obtained that the difference grows as it is expected and for period of $T=64$ days it is only $0.5\%$ of the demanded value $\mathbb{E}[e^{X}]$. One can argue then that, if arbitrage exists, it is at least not so large and can not provide significant profits. Thus, in this discrete time pricing framework, one can take that the log returns are conveniently described with convolutions of the Student's t-distribution. 

Such good approximation of the moment generating function generated solely by using the variance is a consequence of the fact that the most of the probability mass for the Student's t-distribution is within several standard deviations (several $\gamma$ in this case), where the exponential factor $e^x$ is fairly well approximated with a quadratic polynomial. Its explosive effect becomes important at the limits of the integration, which is avoided by the truncation of the distribution. Obviously, the precise relationship between the moment generating function and the variance has more general form
\begin{equation}
    \int_{-x_{\max}}^{x_{\max}}e^x p_T(x) dx = e^{Tf\left(\gamma^2, T\right)},
\end{equation}
where $f$ is some function which encodes the impact of all moments through $\gamma$. We emphasize here that $f$ is a time dependent function, since higher order moments of the convolutions of a distribution do not grow linearly in time in general, as the variance does. Therefore, the distributions we consider here are risk neutral if the following relationship holds
\begin{equation}
r = \mu + f\left(\gamma^2/2,T\right).    
\end{equation}
In this case, for constant rate $r$, the relationship between the stock's growth rate and the unit-period variance will be nonlinear. This could be an engaging topic for future research.

\subsection{Option price sensitivity on the truncation point \label{sec:plateau}}

\begin{figure}[t!]
\includegraphics[width=15cm]{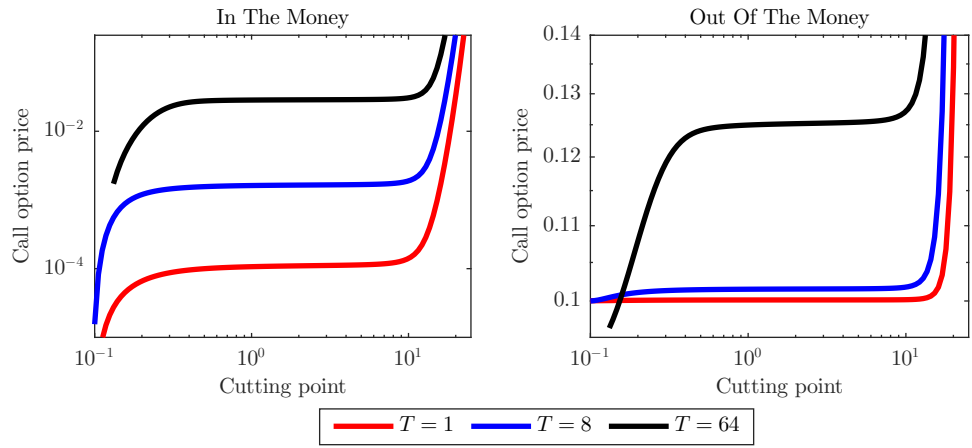}
\caption{Dependence of the option price on the price returns distribution support width $x_{\max}$ for the presented method for call options which are 1 (in red), 8 (in blue) and 64 (in black) days from maturity. The left panel shows an example of in-the money, while at right are given out-of-the money options.}
\label{fig:CallPriceOnWidth}
\end{figure}

As we have seen from the definition of the fair value of call option (\ref{eq:Call_from_price_chng}), using probability density function for log returns which falls off according to power law, would result in infinite option prices. When one chooses to use truncated distributions as a remedy for avoiding the divergence, one is left to determine the truncation point. In order to obtain more insight about it, we have made an analysis of the dependency of the option price on the width of the distribution which is determined from the truncation point. We have found that fitting closest to expiry options with values obtained by our method to those observed from the market, the parameter $\gamma$ has taken values in the range 0.01 and 0.03, so we have opted to make the sensitivity analysis with the middle of that range $\gamma = 0.02$. As an example, in figure \ref{fig:CallPriceOnWidth} is shown the dependence of the option price for strike which is $10\%$ less (at left) or more (at right) than the spot value of a stock taken to be one dollar, on the log return distribution width. The three curves in red, blue and black correspond to options which are 1, 8 and 64 days to maturity and with annual bank rate of $2\%$ which nearly matches the values in the period when the data was studied (end of February and beginning of March, 2018). For better insight, the curves are plotted in logarithmic scales for both axes. As can be seen there is a region of very slow growth, that is narrower for the out-of-the money options. This higher sensitivity is due to the fact that in calculation of such options only the tail part of the distribution contributes, because they will have non zero value at maturity only if the price rises for more than five standard deviations. This region of weak dependence roughly extends from $x_{\max}\approx0.6$ up to $x_{\max} \approx 6$, which corresponds from $M \approx 30$ up to $M \approx 300$ standard deviations for the chosen distribution. A practitioner of this approach could thus take some point in between, for example $x_{\max}=2$ which corresponds to $M = 100$. To obtain a meaningful relationship between the boundaries of the 'plateau' with the statistics of the log returns one could find the total probability which corresponds to the truncated part of the Student's t-distribution. The total probability of occurrence of events from Student's distribution that are apart many standard deviations from the mean can be bounded as in equation (\ref{eq:P_extreme}). Thus, for the lower bound of the plateau, $M \approx 30$, the total probability of such extremes is roughly 1 in 64 000. Since $\gamma = 0.02$ corresponds approximately to price change fluctuation for one day and there are roughly 250 trading days in a year, it means that the events corresponding to the tips of the tails would happen less frequently than once in two and a half centuries. The other border of the price plateau that corresponds for $M \approx 300$, by the same reasoning as above yields probability one in 60 million for the events at the tips of the tails. This translates to situation happening once in 1.2 million years! This analysis suggests that if this power law tail extends long enough, which means that it holds for extremely rare events (up to once in many centuries), the obtained call option prices would be (really) fair. Otherwise, if the power law breaks closer to the body of the distribution, which might mean that the large price changes are impossible, then the option prices obtained with this method would be very sensitive on the point where the distribution is truncated (see the right part of the curves in figure \ref{fig:CallPriceOnWidth}).

For better understanding of this plateau in the table \ref{tab:plateau} are provided the call option prices for a stock with spot price of one dollar, and strikes ten percent more, or less. The parameter of the distribution of the log returns is again $\gamma = 0.02$, and respectively for $M = 100$ standard deviations the truncation point is $x_{\max} = 2$. The left and right boundaries of the plateau were arbitrarily chosen to be $x_{\max, l} = 1$ and $x_{\max, r} = 5$ respectively. The corresponding prices for the left, middle and right boundary points are denoted with $C^{\mathrm{left}}$,  $C^{\mathrm{middle}}$, and $C^{\mathrm{right}}$ respectively. To estimate the inclination of the plateau we have calculated the mismatch between the prices for truncation points at the boundary and the middle as $\Delta C^{\mathrm{left}} = C^{\mathrm{left}} - C^{\mathrm{middle}}$ and $\Delta C^{\mathrm{right}} = C^{\mathrm{right}} - C^{\mathrm{middle}}$. As can be seen for the in-the money options the option prices are rather stable within the plateau, for the three horizons considered: 1, 8, and 64 days. For the out-of-the money options the values are much more sensitive, up to ten percent, which is due to the fact that only the tail part of the distribution is used in calculation of their values and thus the truncation point is rather important.

One can obtain another meaning of the truncation point by observing the price growth corresponding to it. If one chooses to use 100 standard deviations, or $x_{\max} = 2$, the growth factor is $e^{2} \approx 7.4$. This might seem achievable for some stocks for period of several months, but for such stocks one could take higher truncation point, which is a little bit closer to the right boundary of the plateau. For instance, by taking $x_{\max} = 3$, the growth factor is $e^{3} \approx 20$. Although in this case the call option price is more sensitive to the location of the truncation point, growth of of this size for period of a year or two, seems impossible for any asset. We could summarize that, at least for economic growth typical for our age, fair prices of options calculated by using such power law tailed distributions are likely `fair'.

\begin{table}[h!]
  \begin{center}
    \caption{Option price plateau inclination}
   \label{tab:plateau}
    \begin{tabular}{l|r|r|r|r|r} 
    \textbf{Days to expiry (ITM)} & $C^{\mathrm{left}}$ & $C^{\mathrm{middle}}$ & $C^{\mathrm{right}}$ & $\Delta C^{\mathrm{left}}$ & $\Delta C^{\mathrm{right}}$ \\
      \hline
     One day & 0.100 & 0.100 & 0.100 & $-0.000$ & $0.000$ \\
     8 days & 0.101 & 0.102 & 0.102 & $-0.001$ & $0.000$ \\
     64 days & 0.124 & 0.125 & 0.125& $-0.008$ & $0.002$\\
     \hline
\textbf{Days to expiry (OTM)} & $C^{\mathrm{left}}$ & $C^{\mathrm{middle}}$ & $C^{\mathrm{right}}$ & $\Delta C^{\mathrm{left}}$ & $\Delta C^{\mathrm{right}}$ \\     
    \hline
One day & $0.000$ & $0.000$ & $0.000$ & -0.087 & 0.038 \\
     8 days & 0.002 & 0.002 & 0.002 & -0.048 & 0.020\\
     64 days & 0.028 & 0.029 & 0.029 & -0.036 & 0.009\\     
    \end{tabular}
  \end{center}
\end{table}

The existence of a wide enough region of the location of truncation point for which
the call option prices are weakly sensitive is welcomed in option valuation schemes based on truncated distributions. Moreover, the location of the plateau is such that the probability of events that are excluded from consideration in estimation of the option values, is very small. Unfortunately, this has drawbacks as well. In particular, one cannot use a truncated Student's distribution for the log returns in some unit time interval and then strictly apply the independence of consecutive returns for the other, longer periods. The convolution has the property of adding the width of the finite-support probability distributions. In repeated convolutions this results in linear dependence of the width of the distribution on the number of unit periods. Then, the widths of the distributions ranging from one to several dozens of periods cannot fit in the plateau. Thus, either the unit period distribution needs to be too narrow and thus leave the plateau from the left side, or those for longer periods would do the same at the right boundary. For this purpose we choose to truncate \emph{all} distributions at the same point. Clearly, this is in conflict with the statistical independence of consecutive returns. Truncating convolution of two distributions means that, roughly speaking, small and moderate returns are independent but not the large ones. The situations in which extremely large returns are preceded or followed by returns of medium or large size are considered impossible. We believe that abandoning of the statistical independence of log returns of all scales is acceptable, because the extremely large gains or falls are rare. As seen from another perspective, truncating the distributions with different periods at the same point is in fact capping of the log return. Indeed, one might find acceptable to consider that gains exceeding $e^{\gamma M}$ are highly improbable for one or two years, and even nearly impossible for shorter periods. We summarize that this is a favorable trade off for obtaining a family of distributions of log returns, which resembles the observed ones at the tails fairly well and provides option prices weakly sensitive on the truncation point.

\subsection{Practical determination of the distributions of stock returns}

As it was elaborated above, we chose to use the Student's t-distribution for pricing options since it allows analytical treatment to some extent. It has closed form expressions of the Fourier transforms of distributions which are convolutions of it~(\ref{eq:Four_Student_N}). However, inverse Fourier transforms are not easy to obtain. Therefore, we are left with numerical calculation of these inverse Fourier transforms. Instead of making numerical integration as the inverse Fourier Transform is defined, one can use some algorithms for its practical calculation, for example that by \cite{cooley1965algorithm}. This, so called Fast Fourier Transform (FFT) is an efficient approach for calculation of the (discrete) Fourier transform of a sequence $p_n$ and results in a sequence of Fourier coefficients $P_n$. When the original sequence is obtained by sampling some function $p(x)$ then the Discrete Fourier Transform fairly well represents samples of the (continuous) Fourier transform of the function $\mathrm{p} = \mathcal{F}(p(x))$ if the sampling interval is small enough. Thus by applying FFT and its inverse one can obtain samples from the probability density function or from its characteristic function. Next, these samples could be interpolated in order to reach to the desired function or use them directly in numerical routines for respective calculations, like determining values of integrals which correspond to option prices. The usefulness of the FFT is particularly effective in determination of Fourier transforms of irregular sequences (or functions) or those for which analytical results do not exist or are not established yet. For options pricing purpose it means that one can use some empirical probability density of log returns and then by applying FFT and its inverse one can determine probability density for accumulated log returns. We remind that, due to the high sensitivity of the option price when the support of the distribution is outside certain region, one should be careful with the location of the truncation point.

Since algorithms for FFT are rather exploited in different disciplines one can easily find functions for computation of FFT in many programming packages. Then one willing to apply our option pricing procedure needs to implement only numerical routines for calculation of the integrals which can be conveniently represented with sums.

\subsection{A note on Put-Call parity}

In calculation of values of puts, one can apply the well known relationship between the call and put option price for certain stock for same strike price and maturity, which can be found for example in \cite{hull2017fundamentals}. It one of its forms it reads
\begin{equation}
C_T(t) - P_T(t) = S(t) - D K = S(t) - e^{-r(T-t)} K,
\label{eq:Put_Call_definition}
\end{equation}
where $D = e^{-r(T-t)}$ is used as discount factor. If one uses the relations for the call (\ref{eq:Call_from_price_chng}) and put options (\ref{eq:Put_from_price_chng}) and join the integrals will obtain
\begin{eqnarray}
C_T(t) - P_T(t) &=& e^{-r(T-t)}\int_{-\infty}^{\infty} [S_t e^{\mu(T-t) + x} - K] p_T(x) dx \nonumber \\
&=& S_t e^{(\mu - r)(T-t)} \int_{-\infty}^{\infty} e^{x} p_T(x) dx - e^{-r(T-t)} K.
\end{eqnarray}
By applying the approximate no-arbitrage argument (\ref{eq:no_arbitrage}) which holds for distributions with small variance $\gamma \ll 1$, from the last equation one will easily recover the Put-Call parity (\ref{eq:Put_Call_definition}).

\section{Data} \label{sec:data}

In general, the theoretical valuation is focused on European options since they have the simplest form. However, when data for European options is scarce, for empirical analysis one should seek alternatives for testing her or his schemes. In this aspect, as given in \cite{hull2017fundamentals}, one can rely on the fact that American options on non-dividend-paying stocks have the same value as their European variety. Hence, the empirical testing of a pricing scheme of European options can be done by utilizing data on American options. 

For this reason, here we have opted to look at American options freely available data from the Nasdaq's Options Trading Center. The Nasdaq stock market offers daily data free of charge for options of all companies quoted there\footnote{The options data given at the Nasdaq's website correspond to the current values but not historical ones. This means that in order to obtain data for different days one has to visit the website on every such day.}. However, the options for most companies have small sample size and with them credible conclusions can not be reached. Therefore we have restricted the empirical analysis to a selected group of companies that have large market capitalization and whose options are more frequently traded. In addition, we kept in mind for the selected group to be diverse enough in terms of the industrial sectors that are represented. A full list of the used companies as well as the industrial sector they belong to is given in Table~\ref{tab:companies}.

\begin{table}[h]
\centering
\caption{List of studied companies }
\label{tab:companies}
\begin{tabular}{l|l|l}
Symbol & Name & Sector \\
\hline\hline
AAPL & Apple Inc. & Technology   \\
GOOGL & Alphabet Inc. & Technology \\
AMZN & Amazon.com, Inc. & Consumer Services \\ 
MSFT & Microsoft Corporation & Technology \\ 
PEP & Pepsico, Inc. & Consumer Non-Durables \\ 
AMGN & Amgen Inc. & Health Care \\ 
VOD & Vodafone Group Plc & Public Utilities \\ 
TSLA & Tesla, Inc. & Capital Goods \\
\hline
\end{tabular}
\end{table}

For all selected companies we collected the closing information from 28th February until 2nd March for options with maturity up to January 2019. In particular, we collected the maturity of the options, the strike value, the closing bid and ask price of the given day, the daily price change, the volume and the open interest. Since at different stock markets all strike values are not present we chose to use the composite values. Moreover, due to the fluctuating nature of the prices on which the last trade for a certain strike has happened, we chose to take as fair market value the mean of the bid and ask prices. In this way the option price should monotonously change with the strike, although we have observed some exceptions. Finally, for some stocks the call options which are deeply out of the money, there is no bid price, while the ask is very small. We do not take into consideration such cases where the bid value is zero. In the calculations, as a time unit we consider a year consisting of 252 working days. This results in a dataset which covers options whose maturity spans from one day up to 224 days.

\section{Numerical results} \label{sec:results}

The cornerstone of any option valuation framework is the appropriate model for the price dynamics of the underlying financial asset. When one makes a choice for the distributions of the log returns, she is concerned with two slightly confronting demands. First, the chosen probability function needs to match the observed features of the stock price movements. Second, it should be simple enough and allow for analytical treatment for different time scales. The elementary distribution which we decided to use, has been shown to fit to the historical returns rather well. However, for obtaining the distributions for different horizons we must rely on numerical calculations involving (inverse) Fourier transforms, which deliver samples of the distributions instead of closed form formulas. 

In order to have as good as possible sampling we have applied the most dense one we could obtain. Concretely, the numerical calculations were performed on a personal computer and the maximal number of samples from a distribution we could afford\footnote{The Fast Fourier Transform works most efficiently for sequences with length of powers of 2.} was $N_s = 2^{18}$. 
Moreover, when one works with functions which have Fourier transform with infinite support, the distorting phenomenon of aliasing appears. When this phenomenon is not significantly pronounced, the samples of the distribution are perturbed only by a small amount, i.e. the samples obtained from this approach differ only slightly from those of the real distribution. As our focus is put only in the region $(-M\gamma, M\gamma)$ we need samples of the distribution which are equally spaced in that region. After determination of the sample distance as $d = 2M\gamma / N_s$, one obtains the highest frequency in the Fourier transform of the demanded distribution $f_{\max} = 1/2d = N_s / 4M\gamma$. We note that the Fourier spectrum of the distribution of any considered horizon is known in a closed form, because it is simply a power of the Fourier transform of the Student's t-distribution and its samples are thus easily obtained. Once one has the samples of the probability distribution, the option prices are calculated by numerical estimation of the integral (\ref{eq:Call_from_price_chng}) for which we have applied the trapezoidal rule.
 
Our model is solely parameterized by the approximation of the standard deviation $\gamma$ of the probability distribution of log returns. In one approach, this parameter can be estimated from past observations and thus the well known historical volatility can be obtained. In another, it can be inferred from the options on the market as the value which reproduces the market prices best. Here, we have opted for the second scenario, thus taking the `implied volatility' as the measure of standard deviation. More precisely, we do not take the parameter that matches the exact value of the at-the-money option as is usually done. Instead, we choose the optimal parameter to be the one which produces the smallest mean squared error of the differences of logarithms of the theoretical and market option prices for all strikes with certain expiry date for single option. Formally, the mean squared error as a function of $\gamma$ is
\begin{align}
\varepsilon \left( \gamma \right) &= \frac{1}{N_K}\sum_K \left[\log \left(C_T(K, \gamma) \right) - \log\left( C_{\mathrm{market}}(K) \right) \right]^2,
\end{align}
where $C_{\mathrm{market}}(K)$ is the market price of the call on the stock with the same strike $K$ and expiration date and $N_K$ is the number of different strikes for that stock and that maturity. 

This logarithmic error approximately corresponds to the average relative error between theoretical and market prices. The average absolute error in prices is not appropriate here because the out-of-the money options are much cheaper than their in-the money counterparts, and thus the differences are unevenly weighted. We have also determined the optimal parameters for the other two formulas as well. These implied parameters were generated by fitting the market values of the options which have nearest expiry date. The optimal values for all companies which we have studied are provided in Table \ref{tab:imp_parameters}. As pointed out previously, $\gamma$ for the Student's t-distribution has similar value to the volatility of the Black-Scholes-Merton model since both represent standard deviations. However, they are slightly different because they are fit for the whole spectrum of strikes. Note that the parameter for the Borland model is significantly grater since it corresponds to yearly standard deviation, instead of daily which is the unit for the other two formulas. 

\begin{table}[h!]
  \begin{center}
    \caption{Implied distribution parameters by minimizing the mean squared error of log prices of the theoretical option values from the market values.}
    \label{tab:imp_parameters}
    \begin{tabular}{l|l|r|r|r} 
& Company & $\gamma$ & $\sigma_{Bor}$ & $\sigma_{BSM}$ \\
\hline
 28th February & &  &  &  \\
      \hline
     & AAPL & 0.012 & 0.279 & 0.017 \\
     & AMGN & 0.016 & 0.374 & 0.020 \\
	 & AMZN & 0.012 & 0.290 & 0.016 \\     
     & GOOGL & 0.014 & 0.338 & 0.016 \\
	 & MSFT & 0.013 & 0.305 & 0.013 \\     
	 & PEP & 0.010 & 0.223 & 0.011 \\          
     & TSLA & 0.020 & 0.475 & 0.040 \\     
     & VOD & 0.014 & 0.337 & 0.012 \\     
\hline
1st March & &  &  &  \\
      \hline
     & AAPL & 0.015 & 0.405 & 0.025 \\
     & AMGN & 0.019 & 0.502 & 0.027 \\
	 & AMZN & 0.014 & 0.373 & 0.022 \\     
     & GOOGL & 0.017 & 0.463 & 0.021 \\
	 & MSFT & 0.016 & 0.418 & 0.018 \\     
	 & PEP & 0.014 & 0.358 & 0.013 \\          
     & TSLA & 0.021 & 0.592 & 0.051 \\     
     & VOD & 0.018 & 0.471 & 0.014 \\     
     \hline
2nd March & &  &  &  \\
      \hline
     & AAPL & 0.011 & 0.214 & 0.027 \\
     & AMGN & 0.028 & 0.628 & 0.028 \\
	 & AMZN & 0.013 & 0.260 & 0.015 \\     
     & GOOGL & 0.013 & 0.271 & 0.015 \\
	 & MSFT & 0.012 & 0.249 & 0.012 \\     
	 & PEP & 0.012 & 0.243 & 0.012 \\          
     & TSLA & 0.019 & 0.382 & 0.024 \\     
     & VOD & 0.012 & 0.279 & 0.011 \\     
     \hline     
    \end{tabular}
  \end{center}
\end{table}

After the optimal parameters were estimated, the prices for the options with longer maturities were calculated. Again, as an estimate of the accuracy of the pricing algorithm, we used the average difference of the logarithms of the theoretical and market option prices. 
The average errors between the three models and the market values are summarized in Tables~\ref{tab:res_28Feb},~\ref{tab:res_01Mar},~and~\ref{tab:res_02Mar}. Each table corresponds to observations taken from different trading days and for options expiring on different dates. It can be noticed that our model is not able to produce a value for the options with longest maturities. We argue that it is due to the numerical calculations of the Fourier transform, and its resolution is an ongoing research As is shown in \cite{bouchaud2003theory}, the convolutions of Student's t-distribution with three degrees of freedom converge towards the Gaussian with rate $\sqrt{N\log N}$. Namely, the regions of validity of the Gaussian and heavy tails meet approximately at $\sqrt{N\log N} \gamma$, where $\gamma$ is the parameter of the Student's t-distribution. For the longest period considered here, $N=224$, it means that such intersection point is located at approximately 35 standard deviations $\gamma$ from the origin. So, one could use the Gaussian distribution for pricing options with such long period. However, the decision at which period $N$ one could switch from convolutions of the Student's t-distribution to Gaussian, should be examined with a more detailed theoretical analysis. In addition, a comparison of the predictive power of such approach with market option prices with periods within that region is needed as well. Finally one can note that for the other periods, it is evident that for almost all of the considered companies our model performs better than the BSM and Borland models, sometimes resulting in an error that is one order of magnitude smaller.

\begin{table}[h]
\centering
\caption{Mean squared error of the log option prices for data from 28th February}
\label{tab:res_28Feb}
\begin{tabular}{l|l|cccccccc}
        &        & \multicolumn{7}{c}{Days until maturity}     \\
Company & Model  & 2 &7 & 12 & 17 & 21 & 26 & 36 & 224 \\
\hline
\multirow{3}{*}{AAPL}  & Our & \textbf{0.022} & \textbf{0.044} & \textbf{0.031} & \textbf{0.048} & \textbf{0.034} & \textbf{0.043} & \textbf{0.080} & -- \\
& Borland & 0.033 & 0.092 & 0.085 & 0.164 & 0.108 & 0.135 & 0.287 & 0.209    \\
& BSM & 0.071 & 0.047 & 0.052 & 0.102 & 0.121 & 0.160 & 0.091 & \textbf{0.086}    \\
\hline
\multirow{3}{*}{AMZN} & Our & 0.065 & 0.104 & \textbf{0.074} & 0.132 & 0.172 & 0.265 & 0.261 & -- \\
& Borland & 0.082 & 0.108 & 0.263 & 0.065 & 0.132 & 0.063 & 0.155 & \textbf{0.003} \\
& BSM & \textbf{0.026} & \textbf{0.013} & 2.709 & \textbf{0.019} & \textbf{0.269} & \textbf{0.036} & \textbf{0.118} & 0.046    \\
\hline
\multirow{3}{*}{AMGN} & Our & 0.005 & 0.430 & \textbf{0.012} & \textbf{0.017} & \textbf{0.006} & \textbf{0.040} & \textbf{0.020} & -- \\
& Borland & \textbf{0.003} & \textbf{0.122} & 0.214 & 0.415 & 0.313 & 0.682 & 0.584 & 0.607  \\
& BSM & 0.072 & 0.179 & 0.062 & 0.224 & 0.204 & 0.354 & 0.114 & \textbf{0.374}   \\
\hline
\multirow{3}{*}{GOOGL} & Our & \textbf{0.022} & \textbf{0.021} & \textbf{0.009} & \textbf{0.013} & \textbf{0.007} & \textbf{0.012} & \textbf{0.003} & -- \\
& Borland & 0.030 & 0.210 & 0.153 & 0.187 & 0.267 & 0.335 & 0.119 & 0.460 \\
& BSM & 0.026 & 0.025 & 0.021 & 0.065 & 0.037 & 0.052 & 0.019 & \textbf{0.022}    \\
\hline
\multirow{3}{*}{MSFT} & Our & \textbf{0.020} & 0.021 & \textbf{0.026} & \textbf{0.007} & 0.010 & 0.005 & 0.035 & -- \\
& Borland & 0.021 & 0.060 & 0.349 & 0.187 & 0.219 & 0.148 & 0.443 & 0.330 \\
& BSM & 0.020 & \textbf{0.020} & 0.066 & 0.009 & \textbf{0.007} &\textbf{ 0.002} & \textbf{0.011} & \textbf{0.010}    \\
\hline
\multirow{3}{*}{PEP} & Our & 0.025 & \textbf{0.010} & \textbf{0.065} & \textbf{0.233} & \textbf{0.017} & \textbf{0.029} & \textbf{0.151} & -- \\
& Borland & \textbf{0.017} & 0.051 & 0.200 & 0.256 & 0.251 & 0.322 & 0.361 & 0.930 \\
& BSM & 0.105 & 0.059 & 0.383 & 8.151 & 0.055 & 0.057 & 0.269 & \textbf{0.041}    \\
\hline
\multirow{3}{*}{TSLA} & Our & \textbf{0.035} & \textbf{0.034} & \textbf{0.016} & \textbf{0.025} & \textbf{0.021} & \textbf{0.047} & \textbf{0.025} & -- \\
& Borland & 0.039 & 0.264 & 0.274 & 0.276 & 0.317 & 0.213 & 0.729 & \textbf{0.033} \\
& BSM & 0.349 & 1.370 & 0.916 & 1.095 & 1.183 & 0.960 & 1.656 & 0.776    \\
\hline
\multirow{3}{*}{VOD} & Our & 0.061 & 0.044 & 0.050 & 0.038 & 0.021 & 0.044 & 0.038 & -- \\
& Borland & \textbf{0.060} & 0.063 & 0.199 & 0.199 & 0.139 & 0.268 & 0.341 & 1.021 \\
& BSM & 0.063 & \textbf{0.043} & \textbf{0.011} & \textbf{0.012} & \textbf{0.006} & \textbf{0.013} & \textbf{0.003} & \textbf{0.121}

\end{tabular}
\flushleft{\footnotesize Note: Bold denotes lowest error among the three considered formulas.}
\end{table}

\begin{table}[h]
\centering
\caption{Mean squared error of the log option prices for data from 1st March}
\label{tab:res_01Mar}
\begin{tabular}{l|l|cccccccc}
        &        & \multicolumn{7}{c}{Days until maturity}     \\
Company & Model & 1 & 6 & 11 & 16 & 20 & 25 & 35 & 223 \\
\hline
\multirow{3}{*}{AAPL}  & Our & 0.019 & \textbf{0.063} & \textbf{0.059} & \textbf{0.013} & \textbf{0.006} & \textbf{0.014} & \textbf{0.013} & -- \\
& Borland & \textbf{0.016} & 0.605 & 0.505 & 0.666 & 0.521 & 1.119 & 0.874 & \textbf{0.534} \\
& BSM & 0.194 & 0.737 & 0.653 & 0.822 & 0.698 & 1.256 & 0.722 & 0.605 \\
\hline
\multirow{3}{*}{AMZN} & Our & 0.059 & \textbf{0.098} & \textbf{0.060} & \textbf{0.120} & \textbf{0.121} & 0.178 & 0.183 & -- \\
& Borland & \textbf{0.056} & 0.268 & 0.397 & 0.165 & 0.353 & 0.275 & 0.461 & 0.015 \\
& BSM & 0.189 & 0.264 & 0.568 & 0.126 & 0.186 & \textbf{0.167} & \textbf{0.135} & \textbf{0.006} \\
\hline
\multirow{3}{*}{AMGN} & Our & \textbf{0.021} & 0.451 & \textbf{0.014} & \textbf{0.027} & \textbf{0.059} & \textbf{0.068} & \textbf{0.085} & -- \\
& Borland & 0.022 & 0.080 & 0.416 & 0.495 & 0.658 & 0.797 & 0.932 & \textbf{0.598} \\
& BSM & 0.174 & \textbf{0.068} & 0.419 & 0.528 & 0.672 & 0.800 & 0.621 & 0.704 \\
\hline
\multirow{3}{*}{GOOGL} & Our & 0.038 & \textbf{0.087} & \textbf{0.042} & \textbf{0.030} & \textbf{0.022} & \textbf{0.039} & \textbf{0.020} & -- \\
& Borland & \textbf{0.035} & 0.453 & 0.519 & 0.604 & 0.506 & 0.815 & 0.563 & 0.750 \\
& BSM & 0.056 & 0.239 & 0.392 & 0.269 & 0.214 & 0.369 & 0.164 & \textbf{0.352} \\
\hline
\multirow{3}{*}{MSFT} & Our & 0.025 & \textbf{0.036} & \textbf{0.096} & \textbf{0.052} & \textbf{0.036} & \textbf{0.037} & \textbf{0.117} & -- \\
& Borland & 0.024 & 0.248 & 0.767 & 0.517 & 0.536 & 0.559 & 0.932 & 0.576 \\
& BSM & \textbf{0.021} & 0.106 & 0.206 & 0.203 & 0.182 & 0.187 & 0.214 & \textbf{0.137} \\
\hline
\multirow{3}{*}{PEP} & Our & 0.001 & 0.021 & \textbf{0.041} & \textbf{0.111} & 0.091 & 0.129 & \textbf{0.054} & -- \\
& Borland & \textbf{0.001} & 0.205 & 0.549 & 0.874 & 1.083 & 0.980 & 1.407 & 1.824 \\
& BSM & 0.004 & \textbf{0.019} & 0.075 & 6.259 & \textbf{0.082} & \textbf{0.089} & 1.259 & \textbf{0.172} \\
\hline
\multirow{3}{*}{TSLA} & Our & \textbf{0.111} & \textbf{0.054} & \textbf{0.038} & \textbf{0.025} & \textbf{0.027} & \textbf{0.043} & \textbf{0.022} & -- \\
& Borland & 0.125 & 0.529 & 0.730 & 0.645 & 0.709 & 0.681 & 1.279 & \textbf{0.035} \\
& BSM & 1.709 & 2.436 & 2.255 & 2.160 & 2.282 & 2.245 & 3.072 & 1.607 \\
\hline
\multirow{3}{*}{VOD} & Our & 0.021 & 0.043 & 0.098 & 0.133 & 0.146 & 0.242 & 0.240 & -- \\
& Borland & \textbf{0.021} & 0.102 & 0.301 & 0.392 & 0.455 & 0.741 & 0.820 & 0.640 \\
& BSM & 0.021 & \textbf{0.016} & \textbf{0.019} & \textbf{0.039} & \textbf{0.038} & \textbf{0.067} & \textbf{0.063} & \textbf{0.134}

\end{tabular}
\flushleft{\footnotesize Note: Bold denotes lowest error among the three considered formulas.}
\end{table}

\begin{table}[h]
\centering
\caption{Mean squared error of the log option prices for data from 2nd March}
\label{tab:res_02Mar}
\begin{tabular}{l|l|ccccccc}
        &        & \multicolumn{7}{c}{Days until maturity}     \\
Company & Model  & 5                   & 10     & 15     & 19     & 24     & 35     & 222 \\
\hline
\multirow{3}{*}{AAPL}  & Our & \textbf{0.018} & \textbf{0.006} & \textbf{0.015} & \textbf{0.021} & \textbf{0.061} & \textbf{0.033} & -- \\
& Borland & 0.056  & 0.037 & 0.066  & 0.074  & 0.133  & 0.128  & \textbf{0.051}    \\
& BSM & 1.587 & 0.782 & 1.456 & 1.560 & 2.457 & 1.293 & 1.090   \\
\hline
\multirow{3}{*}{AMZN} & Our & 0.048 & \textbf{0.044} & \textbf{0.048} & \textbf{0.073} & \textbf{0.100} & \textbf{0.108} & -- \\
& Borland & 0.114  & 0.116 & 0.105  & 0.158  & 0.115  & 0.128  & 0.012  \\
& BSM & \textbf{0.041} & 0.107 & 0.546 & 0.301 & 0.920 & 0.714 & 0.904  \\
\hline
\multirow{3}{*}{AMGN} & Our & 0.038 & \textbf{0.095} & 0.500 & \textbf{0.267} & 0.865 & 0.738 & -- \\
& Borland & 0.038  & 0.160 & 0.679  & 0.412  & 1.225  & 1.208  & 0.567  \\
& BSM & \textbf{0.022} & 2.577 & \textbf{0.021} & 1.490 & \textbf{0.027} & \textbf{0.172} & \textbf{0.005}   \\
\hline
\multirow{3}{*}{GOOGL} & Our & \textbf{0.028} & \textbf{0.019} & \textbf{0.007} & \textbf{0.005} & \textbf{0.004} & \textbf{0.004} & -- \\
& Borland & 0.080 & 0.094 & 0.048 & 0.124 & 0.123 & 0.054 & 0.242 \\
& BSM & 0.043 & 1.049 & 0.042 & 0.034 & 0.074 & 0.030 & \textbf{0.103}    \\
\hline
\multirow{3}{*}{MSFT} & Our& 0.046 & \textbf{0.011} & \textbf{0.017} & \textbf{0.009} & \textbf{0.006} & \textbf{0.003} & -- \\
& Borland & 0.064  & 0.079 & 0.037 & 0.160 & 0.109 & 0.301 & 0.143  \\
& BSM & 0.034 & 0.142 & 0.042 & 0.033 & 0.010 & 0.079 & \textbf{0.034}    \\
\hline
\multirow{3}{*}{PEP} & Our & 0.010 & \textbf{0.029} & \textbf{0.144} & \textbf{0.073} & \textbf{0.056} & \textbf{0.029} & -- \\
& Borland & \textbf{0.003} & 0.072 & 0.169  & 0.271 & 0.239 & 0.623  & 1.013  \\
& BSM & 0.055 & 0.138 & 6.029 & 0.232 & 0.143 & 0.230 & \textbf{0.684}   \\
\hline
\multirow{3}{*}{TSLA} & Our & 0.040 & \textbf{0.020} & \textbf{0.026} & \textbf{0.025} & \textbf{0.049} & \textbf{0.019} & -- \\
& Borland & \textbf{0.098} & 0.123 & 0.15 & 0.194 & 0.126 & 0.471 & \textbf{0.006} \\
& BSM & 0.057 & 0.406 & 0.055 & 0.069 & 0.042 & 0.089 & 0.074    \\
\hline
\multirow{3}{*}{VOD} & Our & \textbf{0.014} & \textbf{0.012} & \textbf{0.009} & 0.017 & 0.013 & 0.022 & -- \\
& Borland & 0.016 & 0.021 & 0.026 & \textbf{0.044} & \textbf{0.132} & \textbf{0.190} & 0.622  \\
& BSM & 0.014 & 0.012 & 0.015 & 0.014 & 0.005 & 0.011 & \textbf{0.304}
    
\end{tabular}

\flushleft{\footnotesize Note: Bold denotes lowest error among the three considered formulas.}
\end{table}

\begin{figure}[t!]
\includegraphics[width=15cm]{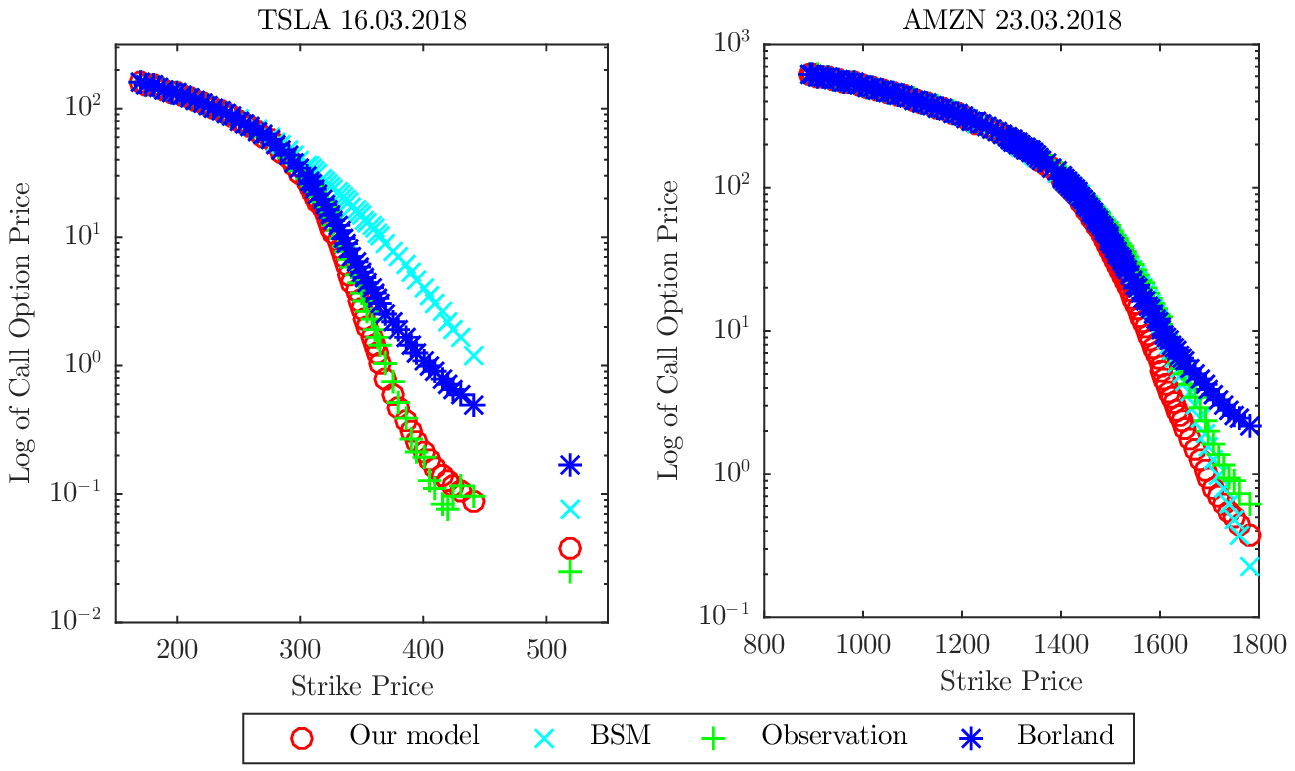}
\caption{ Option prices of the proposed algorithm compared to the Borland formula and Black-Scholes-Merton formula with the respective market values. On the left panel are the results for options prices of Tesla on 1st March 2018, and expiring at 16th March 2018. On the right panel are the corresponding prices for Amazon on 28th February 2018 and expiring on 23rd March 2018.} \label{fig:Opt_Pric_comp}
\end{figure}

\begin{figure}[t!]
\begin{center}
\includegraphics[width=8cm]{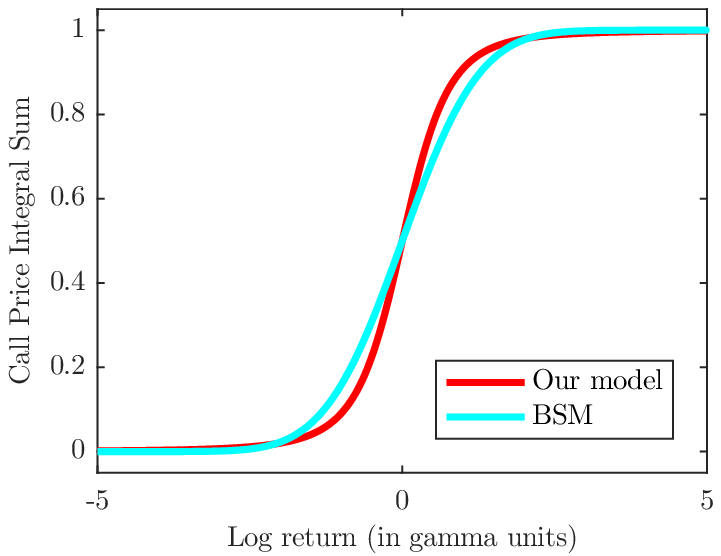}
\caption{Integral sum of the integrals used for calculation of the call option price by using the proposed framework (in red) and the Black-Scholes-Merton approach (in cyan). The unit of the horizontal axis is the distribution parameter $\gamma = \sigma$.} 
\end{center}
\label{fig:Integral_Sum}
\end{figure}

In order to describe the intuition behind these results, in Figure~\ref{fig:Opt_Pric_comp} we illustrate the logarithms of the call option prices of two companies obtained by our approach, together with the Borland and  Black-Scholes-Merton models as a function of the market values which were collected from the Nasdaq web page. The left panel shows a typical case where this approach outperforms the other benchmark models in fitting of the market values. Its predictive power is mainly due to the fact that it traces the curve even in the out-of-the money part of the strike spectrum. When this part is rather wide which means that there are strikes which are much higher than the stock's spot price the other models generally show much weaker performance. When out-of-the money options are not interesting to the market participants, so mainly options which are nearly at the money are traded, the other models can sometimes provide better prediction of the market option prices. One can see one such case in the right panel of Figure~\ref{fig:Opt_Pric_comp}. Another feature that can be observed from the numerical calculations is that the Borland model generally overprices, while the Black-Scholes-Merton formula underprices the out-of-the money options. The weakest side of our approach is the deviation from the market values of the options that are nearly at-the money. From the Figure~\ref{fig:Opt_Pric_comp} it is apparent that the three models produce rather good prediction of the prices of the deeply in-the money options while the deviations are mainly in the other part of the strike spectrum. Such similarity is due to the fact that lion share of price comes from the part of the integral where the body of the distribution lies, which means that the tails do not contribute significantly. To ease understanding of this, we have provided in figure \ref{fig:Integral_Sum} the integral from the lower bound $\log K/S_0 - \mu(T-t)$ up to variable upper bound $x$, as function of the upper bound $x$. Clearly, only few standard deviations $\gamma$ away from the peak of the probability distribution are needed to obtain a good approximation of the price of any deeply in-the money call option.

When considering the deeply out-of-the money options, one should notice that they will be worthless at maturity unless huge growth of the underlying happens. It means that their values are determined from the integral (\ref{eq:Call_from_price_chng}) where only one tail of the distribution contributes. Since exponential tail has much faster decay as compared to the power law peers, the discrepancy between the Black-Scholes-Merton formula and the other two could be understood easily. 

\begin{figure}[t!]
\begin{center}
\includegraphics[width=12cm]{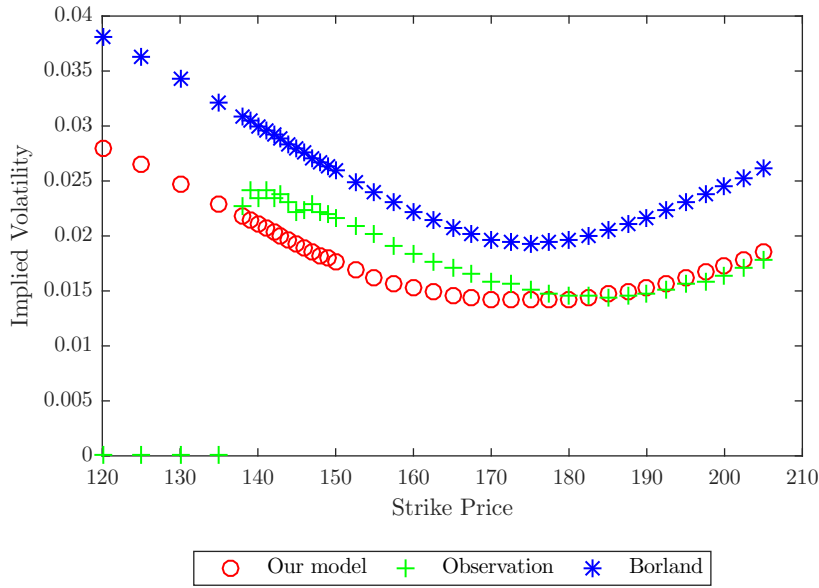}
\caption{Implied volatility curves for our model and that of Borland with those of the market prices. The results shown correspond to the call options of Apple observed at 1 March 2018, which expire three weeks later, at 23 March.} \label{fig:Implied_vol}
\end{center}
\end{figure}

The relevance of an option pricing model is usually determined by verifying its potential to predict the implied volatilities for different strikes. It is assumed to be acceptable if it can generate the volatility curves obtained by determining the volatilities in the Black-Scholes-Merton model that produce the observed market prices. We have not used the implied volatility test in this work because in the data we have studied, for all stocks and nearly for all expiration dates the average of bid and ask market values for some strikes were below the lowest possible option price values. This means that there is no volatility which can result in such fair option price. In fact, such market option prices are below the smallest possible theoretical value which is obtained for zero volatility. This is result of very small bid values that actually represent an arbitrage -- one can immediately exercise the option and make a profit. Regardless of that, we show in figure \ref{fig:Implied_vol} the implied volatilities for the AAPL for options expiring in approximately three weeks. One can notice that the theoretical model of Borland as well as ours produce volatility curves resembling the market one. While the former seems to approach closer to the market determined values for in-the money options, our approach performs better at the out-of-the money section. Also, there are apparently values where the implied volatility for the market price is missing, which are those for which the bid price of the option was below the smallest possible one.

\section{Conclusions} \label{sec:conclusions}

The options pricing scheme proposed in this work relies on the assumption that the future will statistically follow past observations, or that the distribution of log returns is stationary. The truncated Student's t-distribution which appears as a building block of the chain of distributions with different horizons was introduced because it mimics the observed historical returns well, especially at the tails. Options are instruments for which the fairness should be a result of expectation of the future. Without any other insight in the future one could rely on the belief that it will be likely the past and thus use this or some similar probability distribution in calculating the expectations.

The Student's t-distribution and its truncated version besides providing good fit to the observed log returns, were applied in addition due to their simplicity. Sums of variables drawn from such functions do not have closed form of distributions, but have ones for their characteristic functions. It was furthermore obtained that convolutions of such distributions results in models of log returns for different time intervals which are approximately in accordance with the no-arbitrage principle and also support the Put-Call parity relationship. At the end, besides the relative theoretical plausibility, the proposed pricing framework has shown very good accuracy in fitting to the market values of the American options of several companies from different sectors. 


Our motivation was to make a good basis for a pricing framework instead of aiming to design a ready to use pricing formula. Nevertheless, it appeared that the formula produces prices that fit the real data with good accuracy with only \emph{one} parameter. We should emphasize, however, that the observations by \cite{plerou1999scaling} and \cite{amaral2000distribution} of the historical returns have in fact suggested that the distribution is not symmetrical, and moreover the tails of the probability densities of different companies do not fall off exactly as inverse of fourth degree polynomial. This means that one could try to apply a more general Tsallis distribution with appropriate parameter for each company, or even different parameters for the positive and negative returns. The practitioners of theoretical pricing of options might directly apply the proposed procedure, modify it with appropriate tail indices, or even use empirical distributions obtained by their own method and plug them into the pricing framework. The whole procedure in this case would be numerical by using the FFT and its inverse for determination of the distributions of returns for different horizons. The problems we expect to emerge in this case would be related to the implementation of the algorithms for calculation of the Fourier transforms.

A continuous random process for price returns offers distributions of returns for any horizon. This is one of the features which makes the Wiener process which has Gaussian distribution of returns very plausible one. Even though in \cite{borland2002theory} the Student's t-distribution has appeared as a result of a stochastic process with statistical feedback, the related option pricing formula does not coincide with the one presented here. Thus, it remains an open problem whether one could define another stochastic price dynamics process where the probability distributions of log returns would be Student's t-distribution or any truncated version of it. 

As a final remark, we point out that the pricing framework can be fully exploited by determining the option prices for periods to maturity which are are measured to a smaller unit frame than days. In this case one would need to work with very large number of convolutions in order to price options which are several weeks to maturity and may run into numerical obstacles. Another approach for overcoming this problem can be based on finding patterns in change of $\gamma$ for different time intervals and practically use the Student's t-distributions for all horizons. Due to the central limit theorem $\gamma$ would probably grow linearly, as in the Gaussian situation. In order to pursue this way, one needs a huge amount of historical data in order to study the dependence of $\gamma$ on time. Finally, one can apply a combination which would involve Student's t-distribution with different scale parameters for intra day calculations and then make appropriate convolutions for longer periods.

\section*{Funding}

This research was partially supported by the Faculty of Computer Science and Engineering at ``Ss. Cyril and Methodius'' University in Skopje, Macedonia.


\end{document}